\documentclass[fp,twocolumn]{jpsj3}
\usepackage{txfonts}
\usepackage{color}
\usepackage{amsmath}

\title{Structural Phase Transition and Possible Valence Instability of Ce$-4f$ Electron Induced by Pressure in CeCoSi}

\author{Yukihiro Kawamura$^1$, Kakeru Ikeda$^1$, Alisha Nurshafiqah Binti Amat Dalan$^1$, Junichi Hayashi$^1$, Keiki Takeda$^1$, Chihiro Sekine$^1$, Takeshi Matsumura$^2$, 
Jun Gouchi$^3$, Yoshiya Uwatoko$^3$, Takahiro Tomita$^3$, Hiroki Takahashi$^4$, Hiroshi Tanida$^5$}

\inst{$^1$Muroran Institute of Technology, Muroran, Hokkaido 050-8585, Japan\\
$^2$Department of Quantum Matter, AdSE, Hiroshima University, Higashi-Hiroshima, Hiroshima 739-8530, Japan\\
$^3$Institute for Solid State Physics, the University of Tokyo, 5-1-5 Kashiwanoha, Kashiwa, Chiba 277-8581, Japan\\
$^4$College of Humanities and Science, Nihon University, Setagaya, Tokyo 156-8550, Japan\\
$^5$Liberal Arts and Sciences, Toyama Prefectural University, Imizu, Toyama 939-0398, Japan\\}

\abst{X-ray powder diffraction and electrical resistivity measurements were performed on the tetragonal compound CeCoSi under pressure to elucidate
the phase boundary of the pressure-induced structural transition and the change in the 4$f$ electronic state.
The temperature-pressure phase diagram has been determined from the shift of the Bragg peaks and from the anomaly in the resistivity.
The critical pressure, $P_{\rm s}$ $\sim$ 4.9 GPa at 300 K, decreases to $P_{\rm s}$ $\sim$ 3.6 GPa at 10 K.
The decrease of $P_{\rm s}$ is due not only to the decrease in volume of the unit cell but also to an anisotropic shrinkage by cooling.  
When crossing the boundary to the high-pressure phase,
the resistivity shows a significant drop to exhibit a metallic temperature dependence. 
The results of this study strongly suggest that the structural phase transition can be ascribed to valence instability of Ce-$4f$ electron. 

(\today)
}

\begin{document}
\maketitle

\section{Introduction}

$RT_{\rm M}X$ ($R$ = rare earth metal, $T_{\rm M}$ = transition metal, and $X$ = $p$-block element) systems adopt various crystal structures according to their element combinations~\cite{Gupta2015}.
Among them, CeCoSi crystallizes into the tetragonal CeFeSi-type crystal structure (space group No. 129, $P4$/$nmm$, $D^7_{4h}$)~\cite{Bodak1970, Welter1994} that lacks local inversion symmetry at the Ce site.
CeCoSi exhibits an anomalous long-range ordering at an ordering temperature $T_0$ $\sim$12 K~\cite{Tani2019}, followed by an antiferromagnetic (AFM) ordering at the N\'eel temperature $T_{\rm N}$ $\sim$ 9 K~\cite{Tani2019, Chevalier2004, EL2013, Sereni2014} at ambient pressure.
The ordered phases below $T_0$ and $T_{\rm N}$ are defined as Phase II and Phase III, respectively\cite{Tani2019}.
It has been suggested that a possible origin for Phase II is an antiferroquadrupole ordering~\cite{Tani2018, Tani2019, Yatsu2020, Yatsu2020_2, Yatsu2020_3, Mana2021}. 
Odd-parity multipole orderings originating from staggered AFM and antiferroquadrupole orderings have been proposed based on theoretical analyses~\cite{Yatsu2020, Yatsu2020_2}.

A neutron powder diffraction study on CeCoSi shows that the magnetic moments of the two Ce atoms in the unit cell are antiferromagnetically ordered~\cite{SN2020}.
The temperature dependence of the magnetic susceptibility shows that the effective magnetic moment $\mu _{\rm eff}$ is approximately 2.6 $\mu _{\rm B}$, which is close to the free-ion value of 2.54 $\mu _{\rm B}$ for Ce$^{3+}$~\cite{Tani2019}.
In addition, the specific heat studies show that the Sommerfeld coefficient, $\gamma$ $\sim$ 23.9 mJ/molK$^2$, of CeCoSi is as small as that of LaCoSi~\cite{Tani2019,SN2020}. 
These reports strongly suggest that the Ce-$4f$ electrons in CeCoSi at ambient pressure are in a well-localized region in the Doniach phase diagram~\cite{Doniach}.
The Schottky anomaly in specific heat suggests that the first excited state of the crystalline electric field (CEF) is located at $\sim$100 K~\cite{Tani2019, SN2020}.
By applying pressure, the electronic states of CeCoSi can be significantly altered.
An inelastic neutron scattering study suggests that the energy of the first excited level of the CEF increases with the application of 1.5 GPa of pressure~\cite{SN2020}.
$T_{\rm N}$ exhibits a weak pressure dependence below 1.3 GPa, but disappears abruptly at 1.3--1.4 GPa~\cite{EL2013}. 
By contrast, $T_0$ significantly increases under pressure and reaches a maximum of $\sim$40 K at $\sim$1.5 GPa. 
$T_0$ exhibits a rapid decrease above 1.5 GPa, and disappears at $p^*$ $\sim$ 2.2 GPa.

Another intriguing feature of this compound is a structural phase transition induced by pressure at $P_{\rm s}$ $\sim$ 4.9 GPa at 300 K from the low-pressure phase (Phase I) with the tetragonal $P4/nmm$ structure to a high-pressure phase (Phase IV), which is possibly orthorhombic~\cite{YK2020}.
One of the key factors for this structural transition is proposed to be a decrease in the ratio of the tetragonal lattice constants $c/a$.
The $c/a$ of CeCoSi decreases continuously under pressure, which is not observed in the $c/a$ of its isostructural counterparts LaCoSi and PrCoSi, suggesting that Ce-$4f$ electrons play an essential role for the structural transition at $P_{\rm s}$. 
The $c/a$ threshold at $P_{\rm s}$ for CeCoSi is $\sim$1.69~\cite{YK2020}.
The $R$CoSi ($R$ = rare earth) also changes its crystal structure by the lanthanoid contraction; from LaCoSi to TbCoSi the structure is CeFeSi-type, whereas from DyCoSi to LuCoSi the structure changes to an orthorhombic TiNiSi-type~\cite{Dwight1986, Ijjaali1999}.
The fact that the $c/a$ of TbCoSi is $\sim$1.69 implies that the $c/a$ threshold for the structural change in $R$CoSi is $\sim$1.69 and that the structure can be changed either by applying a pressure or replacing the $R$ ion.

The rapid suppression of $T_{\rm N}$, the drastic change of $T_0$, and the emergence of a structural transition by the application of pressure show that the electronic state of the Ce-$4f$ electrons is significantly altered under pressure.
However, the electronic properties in Phases II and III at low pressures and the above structural properties at high pressures have been discussed separately. 
For a comprehensive understanding of the physical properties of CeCoSi, it is necessary to investigate its electronic and structural properties systematically up to high pressures across the $P_{\rm s}$ and determine the temperature-pressure phase diagram in the whole pressure range.

In this study, 
X-ray diffraction (XRD) was performed at pressures up to 6 GPa for $T$ = 6--300 K, and electrical resistivity ($\rho$) measurements were performed at pressures up to 8 GPa for $T$ = 2.5--300 K.
The structural phase transition at $P\rm_{s}$ was observed not only in XRD but also in the resistivity measurements. 
$P\rm_{s}$ decreases with decreasing temperature and reaches 3.6$\pm$0.4 GPa at 10 K,
indicating that the boundary between Phase I and IV is not directly connected to Phases II or III.
The $c/a$ increases with decreasing temperature and decreases with increasing pressure.
Further, the relationship between the low dimensionality of the crystal structure and the decrease in $P\rm_{s}$ at low temperatures and the change in $c/a$ is discussed.
The resistivity exhibits a steep decrease when crossing the boundary from Phase I to IV, after which a simple metallic behavior is observed.  
A strong association between the structural transition and the valence instability is implied.

\section{Experiment}
Single crystals of CeCoSi were prepared using the Ce/Co eutectic flux method, the details of which are presented in the literature~\cite{Tani2019}.
For the XRD experiment, the single crystals were crushed into a powder.
Using the sedimentation method with toluene, a uniform grain of CeCoSi powder was obtained. The powdered sample was pressed into a pellet with a diameter of $\sim$$\phi$0.20 mm and a thickness of $\sim$30 $\mu$m. The pellet was then inserted into a hole in a CuBe gasket with a diameter of $\sim$$\phi$0.25 mm.
Powder XRD under pressure was performed with a helium-gas-driven membrane-type diamond anvil cell (DAC) using synchrotron radiation at the Photon Factory, KEK in Japan. 
The incident beam had a wavelength of 0.6200 $\rm\AA$ and was collimated to a diameter of $\phi$100 $\mu$m. 
The DAC was cooled using the Gifford McMahon (GM) cryostat installed at BL-18C~\cite{Tomita2012}.
The XRD patterns were obtained using an imaging plate with a pixel size of 100 $\mu$m $\times$ 100 $\mu$m.
The pressure in the sample space was determined by the ruby
fluorescence line shifts~\cite{Mao}. 
A mixture of methanol and ethanol at a 4:1 ratio was used as the pressure medium.
The applied pressure was changed at 300 K to realize a hydrostatic pressure.

The resistivity measurements under pressure were performed using a single crystal via the standard four-terminal DC method with a current along the [100] direction. DuPont 4922N silver paste (DuPont de Nemours, Inc., Delaware, USA) was used to attach the gold wire electrode with a diameter of $\phi$20 $\mu$m.
A mixture of Fluorinert FC-70 and FC-77 at a 1:1 ratio was used as the pressure medium.

\section{Results}
\subsection{XRD near the I-IV phase boundary} 

\begin{figure}[ht]
\includegraphics[width=\linewidth]{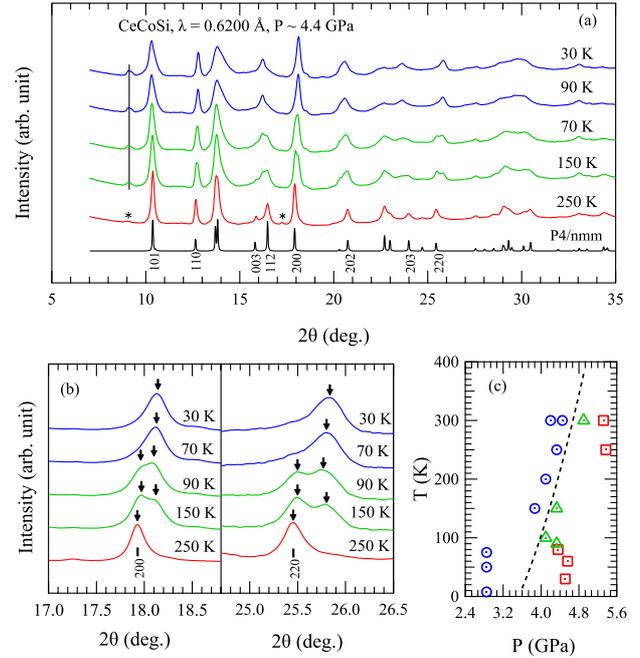}
\caption{(Color online) (a) XRD powder patterns of CeCoSi at various temperatures at $P$ $\sim$ 4.4 GPa. The simulated pattern obtained assuming $P4/nmm$ space group is also plotted. The vertical line and the asterisks indicate the forbidden reflection for $P4/nmm$ and the background anomaly, respectively. See text for detailed explanations. (b) XRD powder patterns for 2$\theta$ $\sim$ 18$^{\circ}$ and 2$\theta$ $\sim$ 25.5$^{\circ}$ are shown in an expanded scale for clarity. (c) Classification of the XRD measurements in the $T$-$P$ space into three categories: Phase I (circle), Phase I+IV (triangle), and Phase IV (square). The dashed line is a guide for the eye, indicating the onset of structural transition.}
\label{XRD}
\end{figure}

Figure \ref{XRD}(a) shows the XRD patterns at $T$ = 250 K, 150 K, 90 K, 70 K, and 30 K at $P$ $\sim$ 4.4 GPa which is just below $P_{\rm s}$ $\sim$ 4.9 GPa at 300 K.
A simulated pattern assuming the $P4/nmm$ space group is also illustrated.
Figure \ref{XRD}(b) shows enlarged views at 2$\theta$ $\sim$ 18$^{\circ}$ and 25.5$^{\circ}$.
As shown in Fig. \ref{XRD}(a), the XRD pattern at 250 K is explained well by assuming the $P4/nmm$ space group. 
The small peaks at 2$\theta$ $\sim$ $9^{\circ}$ and $17^{\circ}$ indicated by the asterisks are ascribed to the signals from the Mylar film of the GM-cryostat window and the CuBe gasket surrounding the sample space, respectively.
At 150 K, additional peaks appear, suggesting a structural transition.
One of the additional peaks is observed at 2$\theta$ $\sim$ $9^{\circ}$ indicated by the vertical line in Fig. \ref{XRD}(a), which corresponds to the 100 forbidden reflection for $P4/nmm$.
Although the background signal from the Mylar also appears at 2$\theta$ $\sim$ 9$^{\circ}$,
the peak at 150 K is stronger than the peak at 250 K.
Other additional peaks are remarkable, at 2$\theta$ $\sim$ 18$^{\circ}$ and 25.5$^{\circ}$.
As shown in Fig. \ref{XRD}(b), a single peak corresponding to the 200 reflection is observed at 250 K, but an additional peak appears at 150 K just above the 200 peak. 
The XRD pattern at 90 K is double peaked similar to the one at 150 K.
At 70 K and 30 K, a single peaked profile is recovered. 
A comparison of the center of the single peak observed at 70 K with that at 250 K shows that it discontinuously shifts to the higher angle. 
The XRD patterns at 70 K and 30 K show that the sample is in Phase IV~\cite{YK2020}.
At an intermediate temperature of 90 K in Fig. \ref{XRD}(b), the XRD pattern consists of the 200 peak observed at 250 K and the single peak observed at 70 K.
Similar behaviors can be seen at 2$\theta$ $\sim$ 25.5$^{\circ}$.
The coexistence of the two phases is observed at intermediate temperatures, indicating that this structural transition is of first-order. This is consistent with our previous report; the XRD pattern at $P\rm_{s}$ $\sim$ 4.9 GPa at 300 K indicates the coexistence of the two phases~\cite{YK2020}.

All of the XRD patterns can be classified into Phase I, Phase IV, and the coexistent state of Phases I and IV (Phase I+IV) in the same manner as the XRD patterns in Fig. \ref{XRD}(a). 
Figure \ref{XRD}(c) illustrates the classification of the measurements in the $T$-$P$ space into the above three categories.
At 300 K, Phase I is maintained up to at least 4.5 GPa; Phase I+IV then emerges at 4.9 GPa, and Phase IV appears above 5.3 GPa. 
With regard to the temperature variation at $P$ $\sim$ 4.4 GPa,
Phase I is maintained down to 200 K; Phase I+IV then emerges at 150 K, and Phase IV appears below 70 K. 
The dashed line in Fig. \ref{XRD}(c) indicates the onset of the structural transition, defined by the phase boundary between Phase I and Phases I+IV. 
This result shows that the boundary between Phase I and IV decreases with decreasing temperature.

\subsection{Electrical resistivity under pressures} 

\begin{figure}[ht] 
\includegraphics[width=\linewidth]{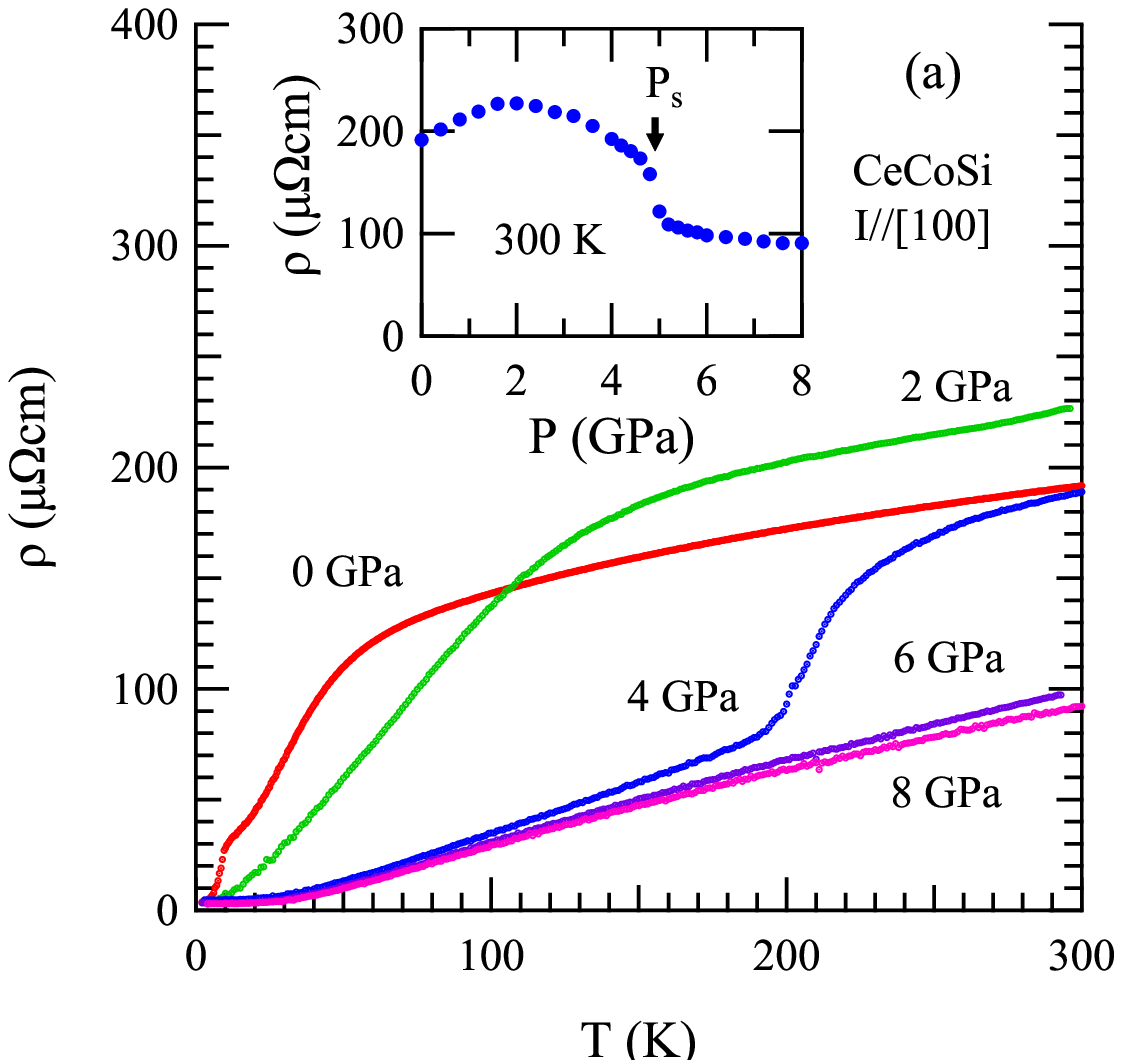}
\includegraphics[width=\linewidth]{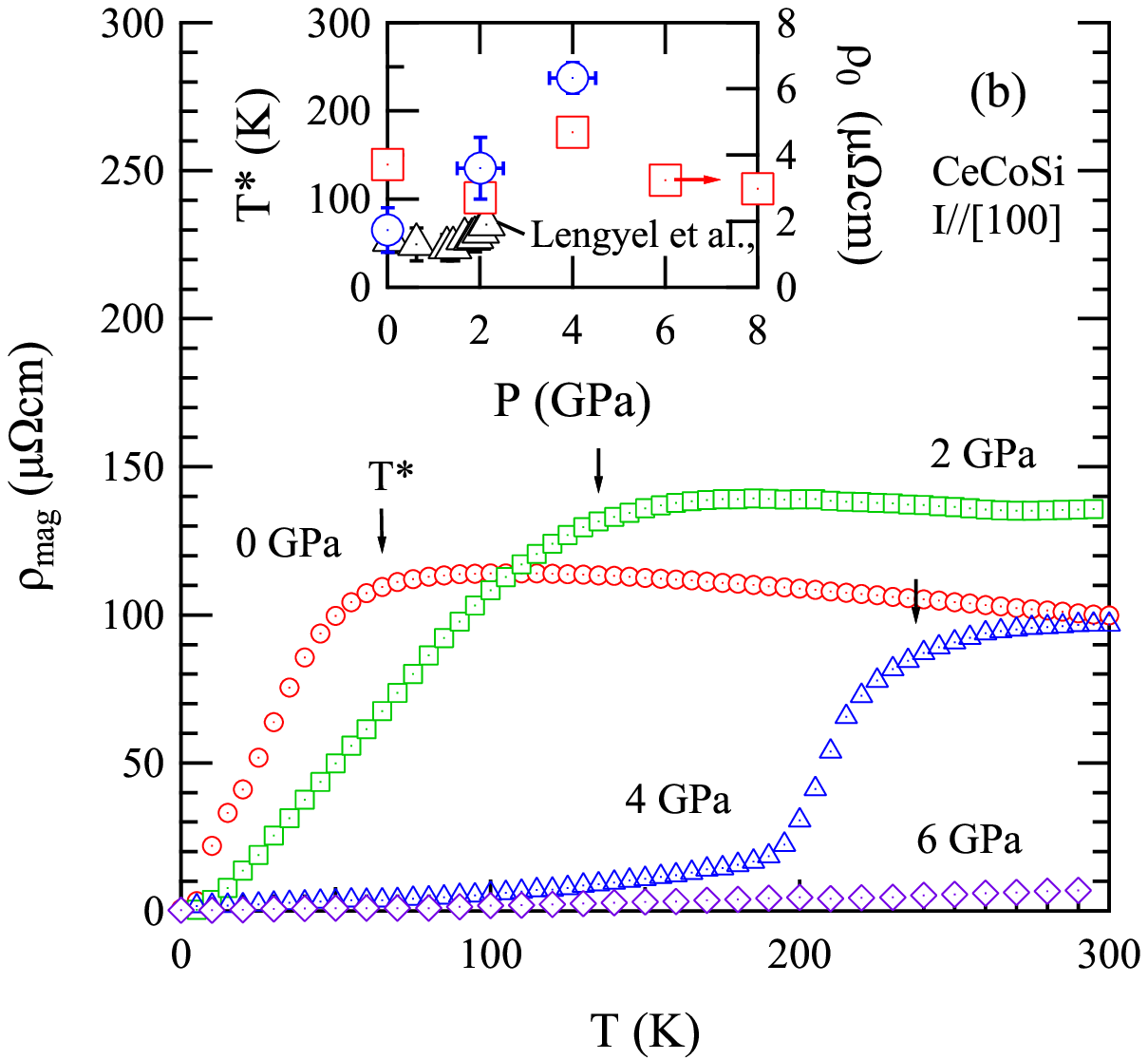}
\caption{(Color online) (a) (Main panel) $T$-dependence of electrical resistivity ($\rho$) of CeCoSi under various pressures. (Inset) Pressure dependence of $\rho$ at 300 K. The arrow indicates the pressure of structural transition $P_{\rm s}$. (b) (Main panel) Temperature dependence of $\rho\rm_{mag}$. Arrows indicate the temperature $T^*$, where $\rho$($T$) exhibits a shoulder. (Inset) Pressure dependence of $T^*$ in this study (left axis, circle) and those taken from the literature (left axis, triangle)~\cite{EL2013} in addition to pressure dependence of the residual resistivity (right axis, square).}
\label{RP}
\end{figure}

\begin{figure}[b]
\begin{center}
\includegraphics[width=\linewidth, clip]{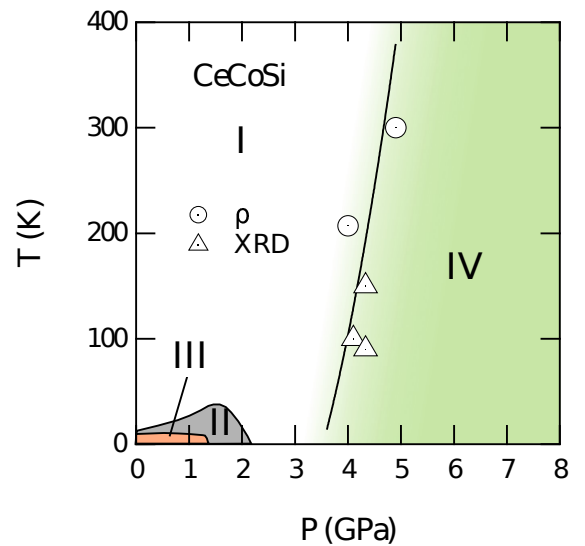}
\caption{(Color online) Temperature-pressure phase diagram of CeCoSi. The boundary of Phases I and IV is obtained from this study. 
The $P\rm{_s}$ defined from resistivity measurement (circle) and Phases I+IV defined from XRD (triangle) are illustrated. Phases II and III from the literature are also illustrated~\cite{Tani2018, Tani2019, EL2013}.}
\label{Phase}
\end{center}
\end{figure}

\begin{figure*}[ht]
\begin{center}
\includegraphics[width=0.33\linewidth]{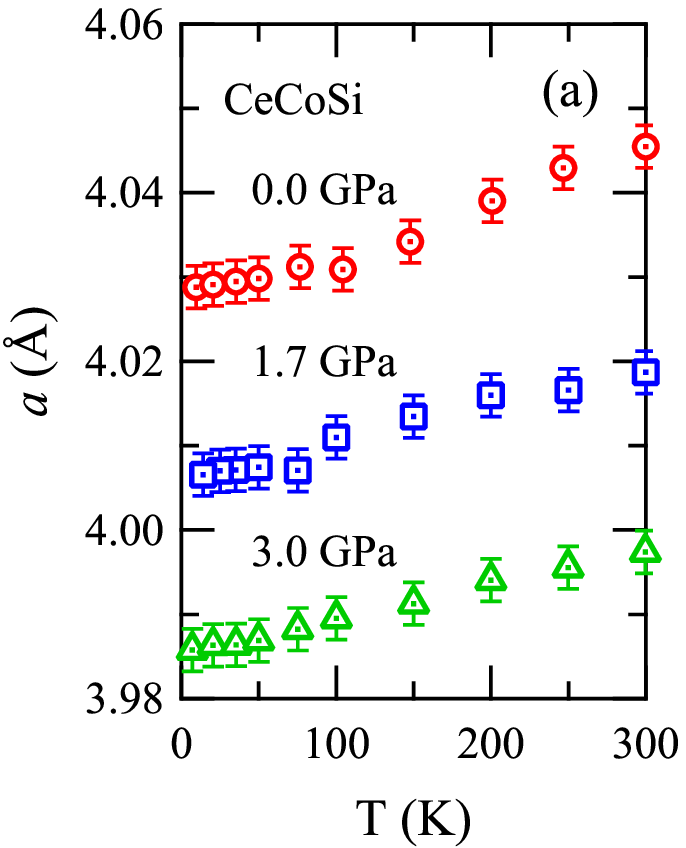}
\includegraphics[width=0.33\linewidth]{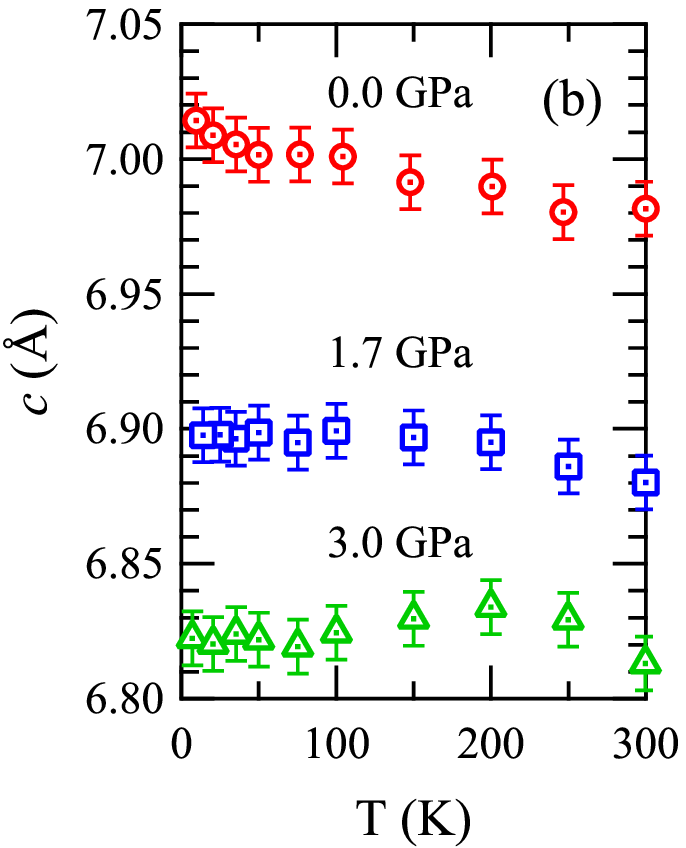}
\includegraphics[width=0.33\linewidth]{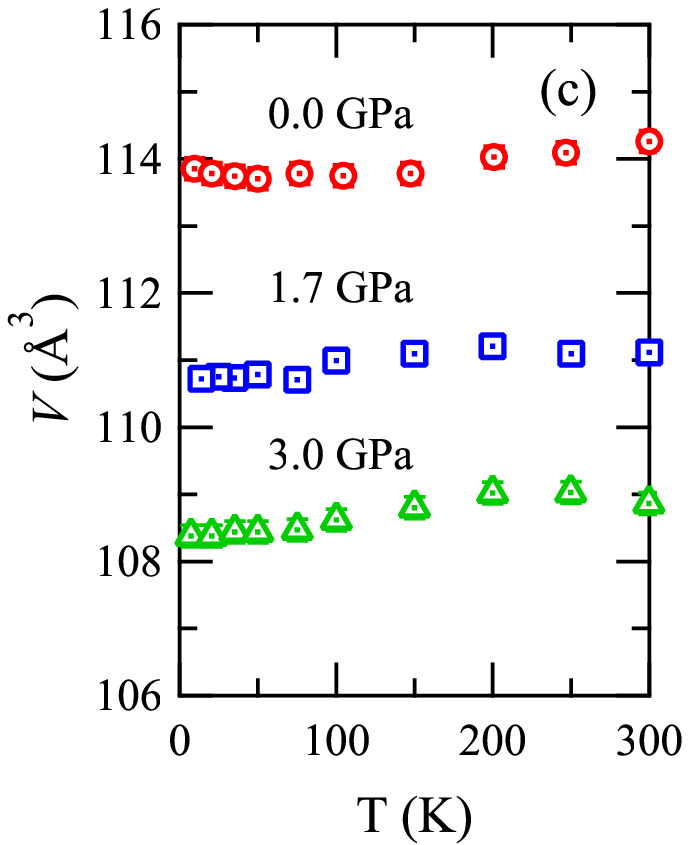}
\includegraphics[width=0.45\linewidth]{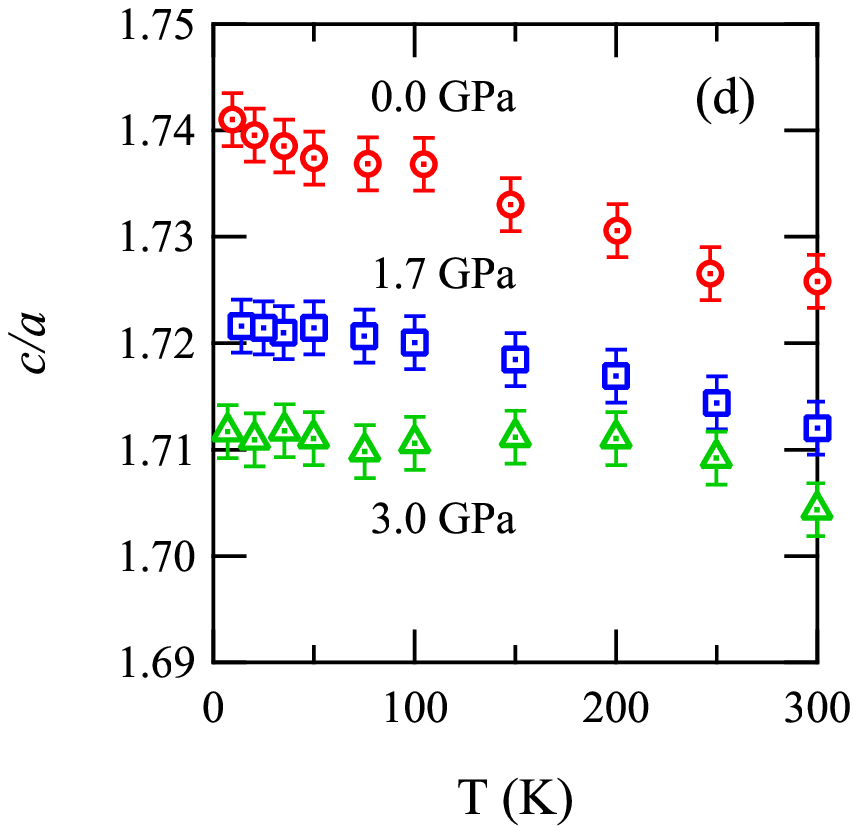}
\includegraphics[width=0.45\linewidth]{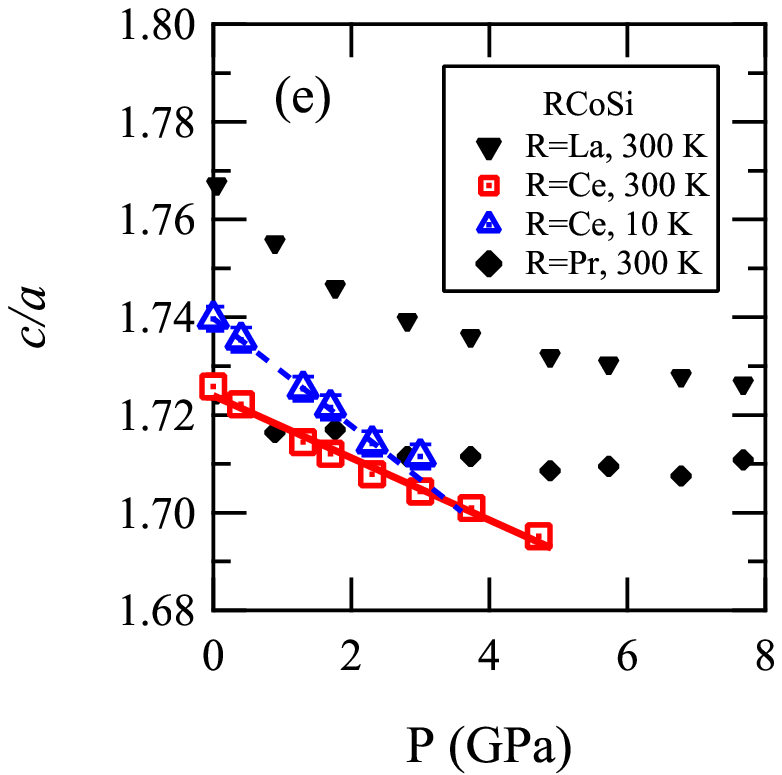}
\caption{(Color online) Temperature dependence of (a) the lattice parameter $a$, (b) the lattice parameter $c$, (c) unit-cell volume $V$, and (d) $c/a$ of CeCoSi at 0.0 GPa (circle), 1.7 GPa (square), and 3.0 GPa (triangle). (e)Pressure dependence of $c/a$ of CeCoSi at 10 K (triangle) and 300 K (square) and those of LaCoSi (inverse triangle) and PrCoSi (diamond) at 300 K~\cite{YK2020}. The straight lines are guides for the eye.}
\label{LC}
\end{center}
\end{figure*}

Figure \ref{RP}(a) shows the temperature dependence of resistivity, $\rho$($T$), for CeCoSi
 under various pressures. 
The inset of Fig. \ref{RP}(a) shows the pressure dependence of $\rho$ at 300 K.  
At 0 GPa, the $\rho$ exhibits a gradual decrease with decreasing $T$ from 300 K, followed by a broad shoulder at $\sim$50 K, a rapid decrease, and a kink at $\sim$9 K. 
This kink corresponds to the antiferromagnetic transition into Phase III. 
The $\rho (T)$ at 0 GPa is consistent with the previous report~\cite{EL2013, Tani2019}. 
The anomaly at $T_0$ is too small to be observed in $\rho(T)$ for both the polycrystalline sample and the single crystal at 0 GPa~\cite{EL2013, Tani2019}.
$\rho (T)$ starts from a slightly higher value than that at 0 GPa, gradually decreases with decreasing $T$, and exhibits a broad shoulder at $\sim$150 K followed by a monotonic decrease down to the lowest temperature. 
Although the $\rho$($T$) data in Ref. 7 exhibit a clear anomaly at $T_0$=12 K at 2.01 GPa, no distinct anomaly was detected in our data at 2 GPa. 
This discrepancy is attributable to a pressure error of the cubic anvil system with an uncertainty of $\pm$0.5 GPa.
The $\rho(T)$ at 2 GPa obtained in this study is considered to correspond to that at 2.4 GPa reported in Ref. 7.
At 4 GPa, the $\rho$ gradually decreases with decreasing $T$ and drops at 210 K, followed by a monotonic decrease almost linearly with $T$, which is reminiscent of a nonmagnetic metal. 
The drop in $\rho$ at 210 K is derived from the structural phase transition from Phase I to when the boundary is crossed, as shown in Fig. \ref{XRD}(c).

As shown in the inset of Fig. \ref{RP}(a),
$\rho$ at 300 K gradually increases with increasing pressure from 0 GPa and exhibits a maximum at 2 GPa, followed by a gradual decrease.
At 4.9 GPa, $\rho$ exhibits a sudden drop by $\sim$40 $\%$.
The discontinuous drop indicates a first order phase transition.
The critical pressure is consistent with $P\rm{_s}$ observed in our previous XRD study at 300 K~\cite{YK2020}. 
As shown in the main panel of Fig. \ref{RP}(a),
$\rho(T)$ monotonically decreases with decreasing $T$ at 6 and 8 GPa.
This is similar to the $\rho(T)$ of LaCoSi at ambient pressure~\cite{Tani2019}.
Additionally, the slope of $\rho(T)$ is almost the same as those at 0 GPa and 2 GPa above the broad shoulder. 
These results suggest that the magnetic scattering due to Ce-$4f$ electrons is strongly suppressed in Phase IV, and the contribution of phonon scattering in Phase IV is almost the same as that of Phase I.
Thus, we assume $\rho(T)$ at 8 GPa as representing the contribution of phonon scattering.
The magnetic part of the resistivity, $\rho\rm_{mag}$, can be treated as $\rho\rm_{mag} = \rho-\rho_{\rm8GPa}$, where $\rho_{\rm8GPa}$ is the resistivity of CeCoSi at 8 GPa.

Figure \ref{RP}(b) illustrates the $T$-dependence of $\rho\rm_{mag}$.
At 0 GPa, $\rho\rm_{mag}$ gradually increases with decreasing $T$ from 300 K, which is consistent with the previous report, where the contribution of phonon scattering was assumed to be the resistivity of LaCoSi~\cite{Tani2019}.
This increase originates from the Kondo scattering of the conduction electrons
by the interaction with the Ce-$4f$ electrons.
The $\rho\rm_{mag}$$(T)$ exhibits a shoulder at $\sim$70 K and rapidly decreases with decreasing $T$.
This decrease is not ascribed to the Kondo coherence but rather to the suppression of magnetic scattering due to the decrease in the scattering channel caused by the reduction of the thermal population of the CEF excited states.
The temperature of this shoulder, $T^*$, increases by the application of pressure as shown in the inset of Fig. \ref{RP}(b).  
The pressure dependence of the residual resistivity $\rho_0$ is also illustrated in the inset of Fig. \ref{RP}(b). 
The $\rho_0$ initially decreases from 0 to 2 GPa but then increases from 2 to 4 GPa.
This behavior can be interpreted as follows.
First, the valence of CeCoSi is most likely to be 3+ at 0 GPa, and $\rho_0$ exhibits a relatively high value due to the magnetic scattering of $4f$ electrons.
At 2 GPa, which may correspond to 2.4 GPa in Ref 7, $\rho_0$ is suppressed since the pressure is above the critical pressure of Phase II.
At 4 GPa, $\rho_0$ increases due to a possible inhomogeneity in the sample occurring in the vicinity of the structural transition.
Above 4 GPa, $\rho_0$ is gradually suppressed due to the decrease of the effect of the structural transition. 
$\rho\rm_{mag}$ ($T$) at 4 GPa shows a large decrease below the temperature of the structural transition.
This strongly suggests that Ce-$4f$ electrons become non-magnetic with the structural transition, which will be discussed in detail in section 4.

Figure \ref{Phase} shows the temperature-pressure phase diagram determined from XRD and $\rho$ measurements.
The regions of Phase II and III taken from the literature are also shown~\cite{Tani2018, Tani2019, EL2013}.
The $P_{\rm s}$ decreases with decreasing $T$. 
The $P_{\rm s}$ $\sim$ 4.9 GPa at 300 K becomes $P_{\rm s}$ = 3.6 $\pm$ 0.4 GPa at 10 K, as shown by the line in Fig. \ref{Phase}. 
The $P_{\rm s}$ $\sim$ 3.6 GPa at 10 K is higher than the critical pressures of Phase II and III.

\subsection{Temperature dependence of lattice parameters under pressure}

Figure \ref{LC}(a) shows the $T$-dependence of the lattice parameter $a$ under pressure.
At 0 GPa, $a$ monotonically decreases with decreasing $T$ from 300 to 100 K, 
followed by a gradual curve below 100 K. 
The $a$ at 300 K is suppressed by applying a pressure of 1.7 GPa and is further suppressed at a pressure of 3.0 GPa.
The decreasing rate $da/dT$ at high temperatures is suppressed from 7.7$\times$10$^{-5}$$\AA$/K at 0.0 GPa to 4.0$\times$10$^{-5}$$\AA$/K at 3.0 GPa, although the $da/dT$ below 100 K at 1.7 and 3.0 GPa is almost the same as that at 0 GPa.
Figure \ref{LC}(b) shows the $T$-dependence of the lattice parameter $c$ under pressure.
At 0 GPa, $c$ increases monotonically with decreasing $T$ from 300 K.
At 1.7 GPa, $c$ increases with decreasing $T$ from 300 to 200 K and becomes almost constant within experimental error below 200 K.
At 3.0 GPa, $c$ increases with decreasing $T$ from 300 to 200 K at a similar rate as those at 0.0 GPa and 1.7 GPa, and exhibits a broad peak at 200 K, followed by a decrease. It is almost constant below 100 K.  
When $a$ and $c$ are compared at 300 K, $c$ is largely suppressed by 2.4 $\%$, whereas $a$ decreases by only 1.2 $\%$. 
It is noteworthy that there is no discontinuity in the lattice parameter and change in the XRD pattern at the boundaries of Phases I and II and Phases II and III under the present experimental accuracy;
no change in the structure was detected in this XRD study where $\Delta a$ and $\Delta c$ exceed 0.005 and 0.02  \AA{}, respectively. 
Figure \ref{LC}(c) shows the $T$-dependence of the unit-cell volume $V$ under pressure.
Although $V$ appears to exhibit a weak decrease with decreasing $T$, the detailed $T$-dependence appears to be non-monotonous, reflecting the different $T$-dependences of $a$ and $c$.

Figure \ref{LC}(d) shows the $T$-dependence of $c/a$ under pressure, which monotonically increases with decreasing $T$ at 0 GPa, reflecting the decrease of $a$ and the increase of $c$.
By the application of pressure, $c/a$ at 300 K is greatly suppressed, which is consistent with our previous XRD results~\cite{YK2020}.
Although $c/a$ increases with decreasing $T$, it decreases with increasing pressure.
At 1.7 and 3.0 GPa, $c/a$ increases with decreasing $T$ from 300 K and exhibits weak $T$-dependence below 100 K, which is associated with the weak $T$-dependences of both $a$ and $c$.
Figure \ref{LC}(e) shows the pressure dependence of $c/a$ at $T$ = 10 and 300 K of CeCoSi. The pressure dependences of the $c/a$ of LaCoSi and PrCoSi at 300 K are also illustrated~\cite{YK2020}. 
The $c/a$ of CeCoSi at 300 K monotonically decreases with increasing pressure at a rate of 6.4$\times$10$^{-3}$ GPa$^{-1}$.
The value of $c/a$ at $P_{\rm s}$(300 K) $\sim$ 4.9 GPa is 1.69.
The $c/a$ at 10 K decreases almost linearly up to 2.3 GPa at a faster rate of 1.1$\times$10$^{-2}$ GPa$^{-1}$.
The $c/a$ at 10 K deviates from the linear pressure dependence at 3.0 GPa, which is a precursor behavior of the structural transition. 
The value of $c/a$ at $P_{\rm s}$(10 K)  $\sim$ 3.6 GPa is 1.70.
The value of $c/a$ at $P_{\rm s}$ is discussed in section 4.
As for LaCoSi and PrCoSi, $c/a$ decreases with increasing pressure at low pressures and then saturates at high pressures without structural transition.
By contrast, the $c/a$ of CeCoSi monotonically decreases with pressure until $P_{\rm s}$, which is considered as reflecting the characteristic $4f$ electronic state close to the valence instability as discussed later.

\begin{figure}[ht!]
\begin{center}
\includegraphics[width=\linewidth]{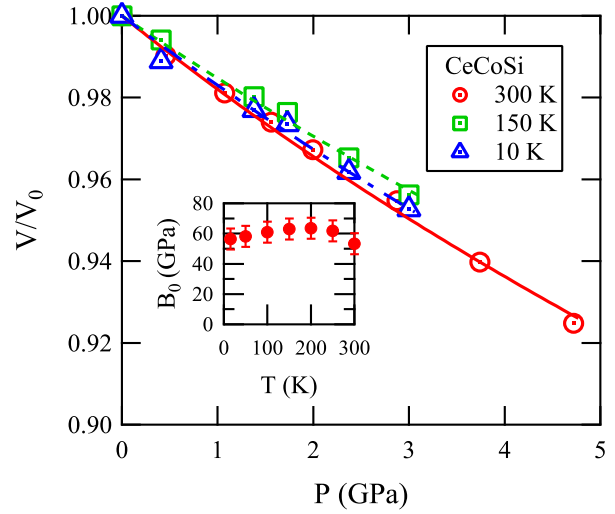}
\caption{(Color online)(Main panel) Pressure dependence of $V/V_0$. The lines indicate the Birch--Murnaghan equation of state fitting; see text for details. (Inset) Temperature dependence of $B_0$.}
\label{B0}
\end{center}
\end{figure}

Figure \ref{B0} illustrates the pressure dependence of the normalized unit-cell volume, $V/V_0$, at 10 K, 150 K, and 300 K, where $V_0$ is the unit-cell volume at ambient pressure.
The $V/V_0$ monotonically decreases with increasing pressure.
The pressure dependence of $V/V_0$ is relatively independent of $T$.
The bulk modulus $B_0$ is determined through least-squares fitting of the following Birch--Murnaghan equation of state~\cite{Birch}:
\begin{equation}
P =\frac{3}{2}B_0 \Biggl[ \biggl(\frac{V}{V_0}\biggr)^{-\frac{7}{3}} - \biggl(\frac{V}{V_0}\biggr)^{-\frac{5}{3}} \Biggr] \Biggl\{1+\frac{3}{4}(B_0'-4)\Biggl[ \biggl(\frac{V}{V_0}\biggr)^{-\frac{2}{3}}-1\Biggr] \Biggr\},
\label{eqB}
\end{equation}
where $B_0'$ is the pressure derivative of $B_0$.
Note that $B_0$ should be interpreted as an approximation for the tetragonal CeFeSi-type structure because cubic symmetry is assumed in this equation.
$B_0'$, representing the curvature of the $P$--$V/V_0$ curve, is sensitive to the pressure range of the fitting. 
We fixed $B_0'$ = 4.0 to compare the $B_0$ values deduced at different $T$. 
The lines in Fig. \ref{B0} are the fitting lines according to Eq. (\ref{eqB}).
The inset of Fig. \ref{B0} shows the $T$-dependence of $B_0$, which increases with decreasing $T$ from 300 to 200 K. 
If we extrapolate the $T$-dependence of $B_0$ to the low temperatures, $B_0$ is expected to reach $\sim$80 GPa at $T$ = 0. 
However, on the contrary, $B_0$ gradually starts to decrease below 200 K, indicating that the crystal becomes softer at low temperatures than at high temperatures.

\section{Discussions}

\subsection{Valence instability for structural transition}

CeCoSi undergoes a structural phase transition at high pressures. From our previous XRD measurements at 300 K, the critical pressure was determined to be $P_{\rm s}$ $\sim$ 4.9 GPa at which the $c/a$ is reduced to $\sim$1.69 from $\sim$1.73 at ambient pressure~\cite{YK2020}. Since the coexistence of the two phases was observed in the vicinity of $P_{\rm s}$, this is the first-order transition.
In this study, we have performed further subsequent XRD measurements at various pressures and temperatures and revealed that $P_{\rm s}$ decreases with decreasing $T$, as summarized in Fig. \ref{Phase}; $P_{\rm s}$ is suppressed down to $P_{\rm s}$ = 3.6$\pm$0.4 GPa at $T$ = 10 K. We have also investigated the electrical resistivity under pressure and found that pronounced anomalies appear at $P_{\rm s}$ determined by the XRD experiments. When passing through $P_{\rm s}$ from Phase I to IV, the electrical resistivity significantly decreases, strongly suggesting that the structural transition involves a drastic change in the Ce-$4f$ electronic state, which will be discussed later. 
The structural instability is not only observed in CeCoSi under pressure but also in the $R$CoSi ($R$ = rare earth) system. Structural instability in the entire $R$CoSi system and a possible candidate of the space group of Phase IV are discussed in the Supplemental Material~\cite{Supple}.

Here, let us discuss the mechanism of the decrease in $P_{\rm s}$ at low temperatures. In general, a crystal lattice contracts at low temperatures.
At first glance, the decrease in $P_{\rm s}$ could simply be regarded as being associated with this contraction by cooling. 
However, as recognized in Fig. \ref{LC}(c), such an effect is far below an order of 1 GPa. 
In addition, as presented in Fig. \ref{LC}(d) and \ref{LC}(e), $c/a$ increases with decreasing $T$, whereas $c/a$ decreases with increasing pressure, indicating that the volume contraction caused by cooling is qualitatively different from that caused by applying pressure.
Thus, the decrease in $P_{\rm s}$ at low temperatures cannot be explained by the volume contraction caused by the cooling.

A key point for understanding the decrease in $P_{\rm s}$ at low temperatures is the low-dimensionality of the crystal structure; Co and Si atoms form two-dimensional $c$-planes, and the Co-Si layers are stacked alternatively with the Ce bilayers along the $c$-axis.
Since the in-plane coupling is supposed to be stronger than the inter-plane coupling due to the presence of the rigid Co-Si layers, the crystal lattice is forced to exhibit an out-of-plane expansion so as to avoid excess volume contraction. 
Consequently, the crystal lattice can contract more easily at low temperatures than at high temperatures when the pressure is applied, because the inter-layer distance becomes large at low temperatures. The lower the temperature, the easier the crystal lattice contracts along the $c$-axis. This is related to the $effectively$ small $B_0$ at low temperatures as mentioned previously.
This scenario provides a qualitative explanation for the larger decrease of $c/a$ at $T$ = 10 K than that at $T$ = 300 K as presented in Fig. \ref{LC}(e), as well as the decrease in $P_{\rm s}$ at low temperatures. 
It is easier to reach the threshold value of $c/a$ for the structural transition at low temperatures than at high temperatures.  
Here, $c/a$ $\sim$ 1.69 at $P_{\rm s}$ $\sim$ 4.9 GPa for $T$ = 300 K, whereas $c/a$ $\sim$ 1.70 at $P_{\rm s}$ $\sim$ 3.6 GPa  for $T$ = 10 K, as seen in Fig. \ref{LC}(e). 
A minor difference in $c/a$ at $P_{\rm s}$ could be related to the effect of the volume contraction caused by cooling as well as a variation in the atomic displacement parameters of the Ce and Si sites along the direction of the $c$-axis. Regardless, the $c/a$ ratio is one of the key parameters for understanding the structural phase transition in CeCoSi at high pressures.

Now we discuss the origin of the structural phase transition at $P_{\rm s}$.
Although the structural transition occurs in CeCoSi under pressure, it is not observed in LaCoSi and PrCoSi. This strongly suggests that the structural transition is related with the Ce-$4f$ electronic states. 
In LaCoSi and PrCoSi, the $c/a$ value is larger than the threshold value of 1.69 even at high pressures, as shown in Fig. \ref{LC}(e). 
In these two compounds, $c/a$ appears to be saturated at high pressures, suggesting that $R$-bilayer cannot shrink even at a high pressure because of the presence of the stable trivalent $R$ ion. The saturated value of $c/a$ appears to depend on the ionic radius of the trivalent $R$ ion. 
However, in the case of CeCoSi, the $c/a$ decreases smoothly down to the threshold value. 
This indicates that the Ce bilayers easily shrink with application of pressure. As a result, the $c/a$ reaches the threshold value, and finally, the structural phase transition takes place at $P_{\rm s}$.
This originates from the valence instability of the Ce ion from Ce$^{3+}$ to Ce$^{4+}$ with a smaller ionic radius. 
We therefore propose that the valence instability is the origin of the structural transition in CeCoSi under pressure. 
Since the $c/a$ of CeCoSi appears to further decrease above 2 GPa without saturating, which is in contrast with the saturating tendency observed in PrCoSi, the Ce valence is suggested to deviate from trivalent to tetravalent at approximately $P$ $\sim$ 2 GPa. Such a valence instability should cause a huge influence on the physical properties as well as on the stability of phase II, which will be discussed in later subsections. 
Direct observation of the valence state of the Ce-$4f$ electron is highly desired. 

\subsection{Valence instability for Phase II}

There are various problems with Phase II to be settled. 
First, the origin of Phase II remains unclear. 
A possibility of multipolar ordering is discussed, but CEF excited states are located far above the ground state. 
Phase II is found to be nonmagnetic, at least from the NMR experiments~\cite{Mana2021}, although $T_0$ does not appear in recent neutron diffraction experiments~\cite{SN2020}. 
The origin of the pressure effects on $T_0$ is also unclear; why $T_0$ is so rapidly enhanced, suppressed, and then disappears with the application of pressure. 
This is in high contrast to the pressure effect on $T_{\rm N}$, which is almost independent of pressure until reaching the critical pressure of the disappearance at $P$ $\sim$ 1.3 GPa. The present experiments, however, shed light on these problems as follows. 
The critical pressure of Phase II is $p^*$ $\sim$ 2.2 GPa, which is roughly 1 GPa lower than $P_{\rm s}$ = 3.6$\pm$0.4 GPa at $T$ = 10 K. 
Experimentally, we have confirmed that Phase I remains up to at least 3.0 GPa. 
In the early electrical resistivity experiments under pressure, there is no clear anomaly of the structural transition up to 2.7 GPa~\cite{EL2013}, which is consistent with the present results of the XRD and electrical resistivity. 
Therefore, Phase II and IV are unambiguously separated by Phase I in the entire temperature range as summarized in Fig. \ref{Phase}. 
Although $p^*$ and $P_{\rm s}$ are separated from each other, the valence instability associated with $P_{\rm s}$ discussed in the previous subsection may affect Phase II especially because it disappears at $p^*$. 
The onset pressure of the valence instability is approximately 2 GPa, above which $T_0$ starts to decrease steeply. 
We thus propose that the steep decrease of $T_0$ may originate from the valence instability associated with the structural transition at $P_{\rm s}$. 
This signifies that the Ce-$4f$ electron may have an important role in the formation of Phase II, whereas whether it is a localized character of the Ce-$4f$ electron remains unclear. 
Considering the tiny released entropy at $T_0$, the Ce-$4f$ electron could influence Phase II via the $c$-$f$ hybridization effect, which could favor forming an odd parity multipole ordering in CeCoSi.

\subsection{Valence instability for electronic properties}

Finally, we discuss the electronic properties of CeCoSi under pressure based on the results of the present electrical resistivity. 
First, at the ambient pressure, the Ce-$4f$ electronic state can be regarded as well-localized; for instance, the magnetic susceptibility exhibits a good Curie-Weiss relation at high temperature, and the Sommerfeld coefficient is as small as that in the nonmagnetic LaCoSi. 
As for the electrical resistivity, the magnetic contribution of the Ce-$4f$ electron to the conduction electron scattering exhibits a weak increase at high temperatures, as presented in Fig. \ref{RP}. This originates from the Kondo scattering. 
The increase of the magnetic contribution itself is also reported in the literature~\cite{Tani2019} and is also often observed in the Ce Kondo lattice system.
A shoulder appears in the electrical resistivity at $T$ $\sim$ 70 K below which the resistivity decreases sharply. In the results of the inelastic neutron scattering experiment, the CEF excited states are separated by an order of 10 meV~\cite{SN2020}. The rapid decrease of the electrical resistivity at low temperature is thus not ascribed to the formation of a heavy electron state but rather to the decrease of the scattering of conduction electrons by localized Ce-$4f$ electrons with CEF splitting. The Kondo scattering due to the CEF excited state is observed in materials such as CeAl$_2$~\cite{Onuki}.

On the other hand, the electrical resistivity is drastically changed when pressure is applied. As shown in the inset of Fig. \ref{RP}(a), the electrical resistivity at $T$ = 300 K increases weakly with increasing pressure up to approximately $P$ $\sim$ 2 GPa, suggesting that the $c$-$f$ hybridization strength becomes large.
Since the effective magnetic moment is reported to be nearly the same up to $P$ $\sim$ 2 GPa even under pressure, the Ce-$4f$ electronic state remains in an almost trivalent state up to $P$ $\sim$ 2 GPa. 
With further increasing pressure, the electrical resistivity begins to decrease, and then exhibits a discontinuous drop at $P_{\rm s}$ $\sim$ 4.9 GPa. 
Above $P_{\rm s}$, the electrical resistivity approaches a constant value of $\rho$ $\sim$ 100 $\mu\Omega$cm as large as that in LaCoSi, indicating that the Ce-$4f$ electronic state becomes nonmagnetic in Phase IV. 
As for the characteristic temperature of $T^*$ where the electrical resistivity shows a broad shoulder, whereas $T^*$ is roughly the same below approximately 1.5 GPa, $T^*$ starts to increase drastically with further increasing pressure, as shown in the inset of Fig. \ref{RP}(b). 
Such a pressure dependence of $T^*$ indicates that the Ce-$4f$ electronic state remains in a Kondo regime up to approximately $T^*$ = 1.5 $\sim$ 2 GPa but begins to change into a valence fluctuating state at higher pressures. 
Since the electrical resistivity is larger in Phase I than in Phase IV, the Ce-$4f$ electronic state remains magnetic up to $P_{\rm s}$; however, the Ce-$4f$ electronic state becomes nonmagnetic in Phase IV.
These characteristic variations of the electrical resistivity under pressure is consistent with the valence state discussed in the previous subsection. 
This is also consistent with the pressure dependence of the CEF excitation under pressure. 
The CEF excitation spectrum is observed until 1.5 GPa, where the excitation energy is enhanced with a marked suppression of the peak intensity of the excitation spectrum~\cite{SN2020}.

At $P$ $\sim$ 2 GPa, the electrical resistivity shows an almost $T$-linear behavior in a wide temperature range below $T^*$, which might reflect an unconventional Ce-$4f$ electronic state originating from the valence instability from a magnetic Ce$^{3+}$ to nonmagnetic Ce$^{4+}$ state. Similar temperature variation of the electrical resistivity was recently reported in Eu-based intermetallic compounds with a valence instability between a magnetic Eu$^{2+}$ to nonmagnetic Eu$^{3+}$ state~\cite{OnukiEu}.

\section{Summary}
In this study, a powder XRD experiment was performed at the temperatures of 6 K $\le$ $T$ $\le$ 300 K and under pressures of 0 GPa $\le$ $P$ $\le$ 6 GPa for CeCoSi.
The XRD results indicate that the structural-transition pressure, $P\rm_{s}$ $\sim$ 4.9 GPa at 300 K, decreases to $P\rm_{s}$ $\sim$ 3.6 GPa at 10 K.
This result indicates that the I to IV phase boundary is not directly connected to that of Phases II or III.
The decrease of $P\rm_{s}$ at low temperatures and unique temperature dependence of a bulk modulus $B_0$ are ascribed to an anisotropic shrinkage of lattice parameters by cooling. 
The electrical resistivity $\rho$ of CeCoSi was also investigated for 2.5 K $\le$ $T$ $\le$ 300 K and 0 GPa $\le$ $P$ $\le$ 8 GPa and indicates a structural transition, consistent with the XRD results.
The $\rho$$(T)$ exhibits a shoulder at $T^*$ $\sim$ 70 K at 0 GPa due to the Kondo scattering and CEF splitting. $T^*$ increases with increasing pressure, which indaicates the enhancement of $c$-$f$ hybridization under pressure.
The resistivity decreases steeply when crossing the boundary from Phase I to IV.
This study suggests that the structural transition is strongly associated with the valence instability. 

\section*{Acknowledgments}
Synchrotron X-ray diffraction was performed at KEK BL-18C, with the approval of the Photon Factory Program Advisory Committee (Proposal Nos. 2019G550 and 2021G545).
A portion of this study was supported by JSPS KAKENHI
Grant Nos. JP17K05546, JP18H03683, JP19K03735, and JP19H00648.


\begin{thebibliography}{99}
\bibitem{Gupta2015} S. Gupta and K. G. Suresh, J. Alloys Compd. {\bf 618}, 562 (2015).
\bibitem{Bodak1970} O. I. Bodak, E. I. Gladyshevskii, and P. I. Kripyakevich, J. Struct. Chem. {\bf 11}, 283 (1970).
\bibitem{Welter1994} R. Welter, G. Venturini, E. Ressouche, and B. Malaman, J. Alloys Compd. {\bf 210}, 279 (1994).
\bibitem{Tani2019} H. Tanida, K. Mitsumoto, Y. Muro, T. Fukuhara, Y. Kawamura, A. Kondo, K. Kindo, Y. Matsumoto, T. Namiki, T. Kuwai, and T. Matsumura, J. Phys. Soc. Jpn. {\bf 88}, 054716 (2019).
\bibitem{Sereni2014} J. G. Sereni, M.G. Berisso, D. Betancourth, V. F. Correa, N. Caroca Canales, C. Geibel, M. G. Berisso, D. Betancourth, and V. F. Correa, Phys. Rev. B {\bf 89}, 035107 (2014).
\bibitem{Chevalier2004}B. Chevalier and S. F. Matar, Phys. Rev. B {\bf 70}, 174408 (2004).
\bibitem{EL2013}E. Lengyel, M. Nicklas, N. C. Canales, and C. Geibel, Phys. Rev. B {\bf 88}, 155137 (2013).
\bibitem{Yatsu2020} M. Yatsushiro and S. Hayami, J. Phys. Soc. Jpn. {\bf 89}, 013703 (2020).
\bibitem{Yatsu2020_2} M. Yatsushiro and S. Hayami, JPS. Conf. Proc. {\bf 30}, 011151 (2020).
\bibitem{Tani2018}H. Tanida, Y. Muro, and T. Matsumura, J. Phys. Soc. Jpn. {\bf 87}, 023705 (2018).
\bibitem{Mana2021}M. Manago, H. Kotegawa, H. Tou, H. Harima, and H. Tanida, J. Phys. Soc. Jpn. {\bf 90}, 023702 (2021).
\bibitem{Yatsu2020_3} M. Yatsushiro and S. Hayami, Phys. Rev. B {\bf 102}, 195147 (2020).
\bibitem{SN2020} S. E. Nikitin, D. G. Franco, J. Kwon, R. Bewley, A. Podlesnyak, A. Hoser, M. M. Koza, C. Geibel, and O. Stockert, Phys. Rev. B {\bf 101}, 214426 (2020).
\bibitem{Doniach}S. Doniach, Physica B {\bf 91}, 231 (1977).
\bibitem{YK2020}Y. Kawamura, H. Tanida, R. Ueda, J. Hayashi, K. Takeda, and C. Sekine, J. Phys. Soc. Jpn. {\bf 89}, 054702 (2020).
\bibitem{Dwight1986} A. E. Dwight, P. P. Vaishnava, C. W. Kimball, and J. L. Matykiewicv, J. Less-Common Met. {\bf 119}, 319 (1986).
\bibitem{Ijjaali1999} I. Ijjaali, R. Welter, G. Venturini, and B. Malaman, J. Alloys Compd. {\bf 292}, 4 (1999).
\bibitem{Tomita2012}T. Tomita, M. Ebata, H. Takahashi, Review of High Pressure Science and Technology {\bf 22}, 222 (2012) [in Japanese].
\bibitem{Mao} H. K. Mao, J. Xu, and P. M. Bell, J. Geophys. Res. Solid Earth {\bf 91} [B5], 4673 (1986).
\bibitem{Birch} F. Birch, Phys. Rev. {\bf 71}, 809 (1947).
\bibitem{Supple}(Supplemental Material)  A discussion of structural instability in $R$CoSi ($R$ = rare earth) between CeFeSi and TiNiSi-type structure and an analysis of XRD pattern in Phase IV are provided online.
\bibitem{Onuki}Y. $\bar{\rm O}$nuki, Y. Furukawa, and T. Komatsubara, J. Phys. Soc. Jpn. {\bf 53}, 2734 (1984).
\bibitem{OnukiEu}Y. $\bar{\rm O}$nuki, M. Hedo, and F. Honda, J. Phys. Soc. Japan {\bf 89}, 102001 (2020).

\end{thebibliography}
\end{document}